\documentclass[a4paper]{article}
\usepackage{amsmath, amsthm, amssymb, amscd}
\usepackage{mathrsfs, eucal,
  amsfonts, latexsym}
\usepackage{epsfig}
\usepackage{graphicx}
\newcommand{\eff}{\mbox{{\sc f}}}
\theoremstyle{plain}
\swapnumbers
\newtheorem{Thm}{Theorem}[section]

\theoremstyle{remark}
\begin{document}

\begin{center}
{\Large Cosmological Particle Creation in States of Low
Energy}
${}$\\[16pt]
{\bf{\large Andreas Degner and Rainer Verch}}
\\[10pt]
Institut f\"ur Theoretische Physik\\
Universit\"at Leipzig\\
Postfach 100 920\\
D-04009 Leipzig, Germany
\end{center}

\begin{abstract}
The recently proposed states of low energy provide a well-motivated class of
reference states for the quantized linear scalar field on cosmological Friedmann-Robertson-Walker
spacetimes. The low energy property of a state is localized close to some value of
the cosmological time coordinate. We present calculations of the relative cosmological
particle production between a state of low energy at early time and another such state
at later time. In an exponentially expanding Universe, we find that the particle production
shows oscillations in the spatial frequency modes. The basis of the method for calculating
the relative particle production is completely rigorous. Approximations are only used at the
level of numerical calculation.
\end{abstract}
\section{Introduction}
The phenomenon of particle production on non-stationary
cosmological spacetimes is one of the early findings
in the theory of quantized fields on curved spacetimes \cite{Parker66}. At its very basics, it can be reduced
to an analogy between the equation of motion for the
mode functions of a quantum field in a time-dependent
(spatially homogeneous) spacetime metric and the 
equation of motion for a harmonic oscillator with time-dependent frequency \cite{Parker66,Fulling}.
It is, however, not easy to make the analogy completely rigorous, for the mode decomposition on which the notion
of particles is based is itself time-dependent.

Quite generally, in a time-dependent background, any concept of particle will itself depend on time, and 
possibly on other data, such as scale parameters.
In more precise terms, for a quantum field on a spacetime with time dependent metric, a concept of particle is tied
to the selection of a reference state, which then depends on such data as a hypersurface of constant time, and other, scale-setting parameters. In the present paper, we will be concerned with the case of a linear, minimally coupled scalar field on Friedmann-Robertson-Walker
(FRW) spacetimes. For this situation, particle creation
with respect to some classes of reference states has
been investigated. Notably, Parker introduced so-called
``adiabatic vacuum states'', designed to minimize particle
creation, as reference states \cite{Parker66,Fulling}.
The precise mathematical definition of this class of 
states is intricate and was first achieved in an investigation by L\"uders and Roberts \cite{LudersRoberts}.
In further work by Junker and Schrohe \cite{JunkerSchrohe},
it was demonstrated that adiabatic vacua are locally
unitarily equivalent to the class of Hadamard states.
Since many results have, over the past decades, confirmed
the view originally put forward by Wald 
\cite{Wald78,WaldQFTCST} that the class of physical states
for linear quantum fields on curved spacetimes should contain
all (quasifree) Hadamard states as building blocks,
this shows that adiabatic vacua qualify as physical states as well. However, the physical interpretation of adiabatic
vacua remains less clear; furthermore, their definition
involves, in the general case, asymptotic series expansions
making them inconvenient to handle and posing difficulties when trying to obtain numerical results
on particle creation \cite{Winitzky}.  

Quite recently, Olbermann \cite{Olbermann} has introduced
a new class of reference states for the linear scalar
field on FRW spacetimes, called {\it states of low energy}
(SLEs). These states are spatially homogeneous (with
respect to the spatial isometry group of FRW spacetimes)
and such that that they minimize the time-integral of
the energy-density for a given weighting function.
For the moment, let us describe this in more detail
as follows (the exact definition will be given in Sec.\ 3).
Suppose that $t$ is the cosmological time coordinate in
an FRW spacetime, and let $f(t)$ denote a non-negative,
compactly supported, smooth function of cosmological
time. For a homogeneous state of the quantized linear
scalar field on FRW spacetime, denoted by its expectation
functional $\langle\,.\,\rangle_\omega$ where 
$\omega$ is a label for the state, we write
$\langle :\boldsymbol{\varrho}(t): \rangle_\omega$ for
the (renormalized) expected energy density in that state.
As the state is homogeneous, this quantity is only
dependent on time $t$. The time-integral of the
expected renormalized energy density, weighted with respect to
$f$, is
$$ \varrho_\omega[f] = \int \langle :\boldsymbol{\varrho}:(t)
 \rangle_\omega\,f(t) \,dt \,.$$
Then a state $\langle\,.\,\rangle_{\omega_{\rm sle}}$ is called
an {\it SLE (state of low energy) with respect to the
weighting function $f$} if the state minimizes the
expression
$\varrho_\omega[f]$ among all homogeneous states  
$\langle\,.\,\rangle_\omega$.
Note that the SLE property depends on the choice of
weighting function $f$ and that it is therefore a concept
that is local in time. Olbermann has shown that SLEs
are Hadamard states, and that adiabatic vacua are
approximations of SLEs. 

If a state $\langle\,.\,\rangle_\omega$ were an SLE for
the case $f = \delta_{t_0}$, the delta-distribution
concentrated at cosmological time $t$, it would correspond to a state minimizing the (homogeneous) energy density
at time $t_0$. This is reminiscent of the concept of
an instantaneous vacuum state at time $t_0$, and the
interpretation of SLEs is that of approximate (or,
by weighting with respect to $f$, mollified) 
instantaneous vacua at time $t_0$.  Instantaneous vacua,
however, aren't Hadamard states, and they are not even
(locally) unitarily equivalent to Hadamard states
\cite{Fulling79,JunkerRMP}, thus
disqualifying them from the class of physical states.
The weighting, or mollifying, of the energy density
by integrating in time against smooth weighting functions $f$ in
defining SLEs is therefore needed in order to ensure that
SLEs are physical states. 

Nevertheless, the interpretation as homogeneous, 
energy minimizing states close to an instant
of cosmological time makes SLEs natural candidates
for reference states with respect to which particle
concepts can be defined and particle creation in
FRW spacetimes can be calculated. This is the topic
of the present work. More specifically, we consider
two smooth weighting functions, $f_1$ and $f_2$,
peaked at cosmological time parameters $t_1$ and
$t_2$, respectively. We will envisage the situation
that $t_1$ corresponds to an early cosmological time
and $t_2 > t_1$ a later cosmological time.
Denoting by $\langle \,.\,\rangle_{\omega(1)}$ and
$\langle\,.\,\rangle_{\omega(2)}$ the corresponding
SLEs, we will determine the particle content per
(fixed)
frequency mode $k$ (arising from the spatial symmetry) of
the states with respect to each other; this quantity is given by the
modulus squared $|\beta_k|^2$ of the Bogoliubov coefficients, see Sec. 4. The
numerical calculation, however, turns out to be involved,
and therefore we limit ourselves in this work to 
a special case which simplifies the calculations
significantly. However, we think that the principal
qualitative aspects of the results of our calculations won't depend very
critically on these choices. Our specialized assumptions
are: (i) A closed (spatially compact) FRW spacetime,
(ii) an exponentially increasing scale-factor $a(t) =
e^{Ht}$ ($H > 0$), (iii) a very small mass parameter $m > 0$,
fine-tuned with respect to the Hubble-parameter $H$,
in the linear field equation of the quantized scalar field.
Moreover, we approximate the weighting functions $f_1$ and $f_2$
by Gaussians peaked at $t_1$ and $t_2$ with small characteristic widths
$\epsilon$ and $\delta$, respectively. Then $|\beta_k|^2$ will be
calculated numerically and represented graphically for certain choices of
$t_1,t_2,\epsilon$ and $\delta$. 

The most interesting result is that $|\beta_k|^2$ shows oscillations in $k$ when
$t_2 -t_1$ is different from 0 and this effect increases with growing $t_2 - t_1$.
A somewhat related observation to the effect that the particle number per mode can
decrease in time appears in \cite{HuKM}; however, the reference states used in
\cite{HuKM} are instantaneous vacua.

It is worth pointing out that our appraoch is completely rigorous, with approximations
entering only at the level of numerical calculation. While the simplifications 
(i), (ii) and (iii) made in order to facilitate the numerical calculations are not physically realistic,
it is, as mentioned, to be expected that the basic findings of our calculations,
especially the said oscillatory behaviour of $|\beta_k|^2$, will qualitatively also
occur in physically more realistic situations. This could have some implications with
regard to observations in cosmology. 

This article is organized as follows. The quantization of the linear (minimally coupled) scalar field
on FRW spacetimes will be summarized in Sec.\ 2, together with the definition of the 
renormalized stress-energy tensor. States of low energy will be discussed in Sec.\ 3.
In Sec.\ 4 we consider Bogoliubov transformations between two SLEs and the associated
notion of particle creation. The numerical calculations will be presented in Sec.\ 5.
Summary and Outlook are given in the final Sec.\ 6.

\section{Quantum field on FRW spacetimes and renormalized stress-energy tensor}
In this section, we summarize the quantization of the linear scalar Klein-Gordon field
on FRW spacetimes, and the definition of the renormalized stress-energy tensor. This serves mainly to
make the present text as self-contained as possible. 
General references for the material in this section are
\cite{Wald78,WaldGR,WaldQFTCST,Fulling}.

A spacetime $(M,g)$, where $M$ denotes the spacetime manifold and $g$ the Lorentzian
metric, is of FRW type if it is of the form
\begin{enumerate}
\item $M = \mathbb{R} \times \Sigma$
\item $ds^2_g = dt^2 - a(t)^2 h_{ij}d{\bf x}^i d{\bf x}^j$ \item $(\Sigma,h)$ is a 3-dimensional manifold with Riemannian metric of constant curvature
which is either equal to 0 or normalized to $\pm 1$.
(We write usually $x$ for elements in $M$, identified 
also as $(t,{\bf x})$ with $t \in \mathbb{R}$,
${\bf x} \in \Sigma$.)
\item $a : \mathbb{R} \to \mathbb{R}$  is a $C^\infty$ function taking strictly positive values,
called the scale factor.
\end{enumerate}
The classical linear, minimally coupled scalar field (Klein-Gordon field) $\varphi$ on $(M,g)$ fulfills
the field equation 
$$ (\Box_g + m^2) \varphi = 0$$
with smooth $\varphi: M \to \mathbb{R}$. The constant parameter $m \ge 0$ is called the mass term of the
field equation and $\Box_g$ is the d'Alembertian of $(M,g)$ which in the present case takes the form
\begin{equation} \label{KG-FRW}
 \Box_g + m^2 = \frac{\partial^2}{\partial t^2} + 3 H(t) \frac{\partial}{\partial t} -
    \frac{1}{a(t)^2} \Delta_h + m^2 \,.
\end{equation}
Here, $H(t) = (\frac{d}{dt}a(t))/a(t)$ is the Hubble function, and $\Delta_h$ denotes the Laplace
operator on $(\Sigma,h)$.
An FRW spacetime $(M,g)$ is a globally hyperbolic spacetime and therefore there exists a unique pair,
$E^\pm$, of advanced/retarded Green's functions for the Klein-Gordon operator $\Box_g + m^2$, for
any fixed $m \ge 0$ \cite{Dimock,BGP}. They are characterized as continuous linear maps 
$E^\pm : C_0^\infty(M,\mathbb{R}) \to C^\infty(M,\mathbb{R})$ with
$E^\pm (\Box_g + m^2) \eff = \eff = (\Box_g + m^2)E^\pm \eff$ for all $\eff \in C_0^\infty(M,\mathbb{R})$,
and with ${\rm supp}(E^\pm \eff) \subset J^\pm({\rm supp}\,\eff)$ for all $\eff \in C_0^\infty(M,\mathbb{R})$,
where $J^\pm(G)$ denotes the future($+$)/past($-$) causal set of a subset $G \subset M$.
The difference $E = E^+ - E^-$ is often called causal propagator of the linear scalar field on
$(M,g)$. We write 
$$ \mathcal{E}(\eff_1,\eff_2) = \int_{M} \eff_1(x) (E\eff_2)(x)\, d\mu_g(x) \quad (\eff_1,\eff_2 \in C_0^\infty(M,\mathbb{R}))$$
for the associated bilinear form, where $d\mu_g(x) = \sqrt{|{\rm det}(g_{\mu\nu})|}d^4x$ is the metric-induced
volume form on $M$. One can show that $\mathcal{E}$ is antisymmetric, i.e.\\ $\mathcal{E}(\eff_1,\eff_2) = -
\mathcal{E}(\eff_2,\eff_1)$.

The linear scalar field on $(M,g)$ can then be quantized as follows. One defines an abstract $*$-algebra 
$\mathcal{F}(M,g)$ with algebraic unit element ${\bf 1}$ as being generated by ${\bf 1}$ and a family
of elements $\phi(\eff)$, $\eff \in C_0^\infty(M,\mathbb{R})$, with the relations
\\[4pt]
${}$ \quad $\phi(\eff_1)\phi(\eff_2) - \phi(\eff_2)\phi(\eff_1) = \frac{i}{2}\mathcal{E}(\eff_1,\eff_2)$,
\\[2pt]
${}$ \quad $\phi(\eff)^* = \phi(\eff)$,
\\[2pt]
${}$ \quad $\phi((\Box_g + m^2)\eff) = 0$
\\[2pt]
for all $\eff,\eff_1,\eff_2 \in C_0^\infty(M,\mathbb{R})$.
\\[4pt]
One can show that such an algebra exists (nontrivially) and that it is unique up to (natural) isomorphisms
(cf.\ e.g.\ \cite{BFV} and lit.\ cited there). The generating elements $\phi(\eff)$ can be viewed as ``abstract field operators''
for the quantized linear scalar field on $(M,g)$. The $\phi(\eff)$ can be turned into operators 
acting in a Hilbert space upon considering Hilbert space representations of $\mathcal{F}(M,g)$.
Here, we are interested in Hilbert space representations induced by states on $\mathcal{F}(M,g)$.
Recall that a state on $\mathcal{F}(M,g)$ is a linear functional $\omega : \mathcal{F}(M,g) \to 
\mathbb{C}$ which is positve, i.e.\ it fulfills $\omega(X^*X) \ge 0$ for all $X \in \mathcal{F}(M,g)$.
(Note that $X$ is in $\mathcal{F}(M,g)$ if it is a polynomial built out of 
${\bf 1}$ and finitely many $\phi(\eff_1),\ldots,\phi(\eff_N)$.) Moreover, it is required that $\omega$ be 
continuous, that is, for each $n \in \mathbb{N}$ the map
$\eff_1 \otimes \cdots \otimes \eff_n \mapsto \omega(\phi(\eff_1) \cdots \phi(\eff_n))$ extends to a distribution
in $(C_0^\infty(M^n,\mathbb{R}))'$. It is also common to denote a state by its expectation value
functional, $\langle X \rangle_\omega = \omega(X)$, and we shall often adopt this notation.

There are states $\omega$ entirely determined by their 2-point function
$$ \mathcal{W}_2^\omega(\eff_1,\eff_2) = \omega(\phi(\eff_1)\phi(\eff_2)) = \langle \phi(\eff_1)\phi(\eff_2) \rangle_\omega
\quad (\eff_1,\eff_2 \in C_0^\infty(M,\mathbb{R})) $$
by requiring the relations $\langle \phi(\eff_1) \cdots \phi(\eff_{2n +1}) \rangle_\omega = 0$ and 
$$
  \left. \frac{d^{2n}}{d \lambda^{2n}}\right|_{\lambda = 0} \langle {\rm e}^{i \lambda \phi(\eff)} \rangle_\omega
   =  \left. \frac{d^{2n}}{d \lambda^{2n}}\right|_{\lambda = 0} {\rm e}^{- \lambda \mathcal{W}_2^\omega(\eff,\eff)/2}
   \quad (\eff \in C_0^\infty(M,\mathbb{R}))
   $$
for all $n \in \mathbb{N}$; the left hand side of this equation is to be read as the expectation value of the
polynomial in $\phi(\eff)$ that results from formally differentiating
${\rm e}^{i\lambda \phi(\eff)}$ and setting $\lambda$ equal to $0$.

A state $\omega$ on $\mathcal{F}(M,g)$ induces a $*$-representation $\pi_\omega$ of $\mathcal{F}(M,g)$ on a
dense domain $\mathcal{D}_\omega$ in a Hilbert space $\mathcal{H}_\omega$ together with
a canonical unit vector $\Omega_\omega \in \mathcal{D}_\omega$. The collection of $(\mathcal{H}_\omega,\mathcal{D}_\omega,
\pi_\omega,\Omega_\omega)$ is called GNS-representation, or Wightman-representation of $\omega$ (cf.\ \cite{StreaterWightman})
and is characterized by the properties that $\langle X \rangle_\omega = \langle \Omega_\omega, \pi_\omega(X) \Omega_\omega \rangle$
for all $X \in \mathcal{F}(M,g)$, $\pi_\omega(X^*) \chi = \pi_\omega(X)^* \chi$ for all $\chi \in \mathcal{D}_\omega$,
and $\mathcal{D}_\omega = \{\pi_\omega(X)\Omega_\omega : X \in \mathcal{F}(M,g)\}$. Then the represented abstract field operators 
$\phi(f)$ become unbounded operators 
\begin{equation}\label{fieldop1}
 \Phi_\omega(f) = \pi_\omega(\phi(f)) 
\end{equation}
defined on the domain $\mathcal{D}_\omega$ in the representation Hilbert space $\mathcal{H}_\omega$.

For a quasifree state $\omega$, there is a one-particle Hilbert space $\mathcal{H}_\omega^{(1)}$ together with a real-linear
map $K_\omega : C_0^\infty(M,\mathbb{R}) \to \mathcal{H}_\omega^{(1)}$ so that 
$\langle K_\omega(\eff_1),K_\omega(\eff_2) \rangle = \mathcal{W}_2^\omega(\eff_1,\eff_2)$. One can then show that
$\mathcal{H}_\omega = F_+(\mathcal{H}_\omega^{(1)})$, the bosonic Fock-space over the one-particle space
$\mathcal{H}_\omega^{(1)}$, and that 
\begin{equation} \label{fieldop2}
\pi_\omega(\phi(\eff)) = a(K_\omega(\eff)) + a^+(K_\omega(\eff)) 
\end{equation}
where $a(\,.\,)$ and $a^*(\,.\,)$ are the usual annihilation and creation operators on
$F_+(\mathcal{H}_\omega^{(1)})$. Furthermore, $\Omega_\omega = (1,0,0,\ldots)$ is the 
Fock-vacuum-vector and thus it is convenient to write $|0\rangle_\omega = \Omega_\omega$
for the GNS-vector of a quasifree states $\omega$. Moreover, a quasifree state $\omega$ is {\it pure} if
the range of $K_\omega$ is dense in $\mathcal{H}_\omega^{(1)}$. This is equivalent to saying that
$\omega$ cannot be written as convex combination of several different states. 
    
On each FRW spacetime $(M = \mathbb{R} \times \Sigma,g)$ there acts the spatial symmetry group
$G_\Sigma$. It acts only in the $\Sigma$-part of $M$, so that for each $\gamma \in G_\Sigma$ one has
$\gamma(t,{\bf x}) = (t,\gamma_\Sigma({\bf x}))$ for all $t \in \mathbb{R}$, ${\bf x} \in \Sigma$ with respect to the
$\mathbb{R} \times \Sigma$-splitting of $M$, with an isometry $\gamma_\Sigma$ of $(\Sigma,h)$.\footnote{Strictly
speaking, $G_\Sigma = \{{\rm id}_\mathbb{R} \times \gamma_\Sigma : \gamma_\Sigma \in {\rm Iso}^+(\Sigma,h) \}$
where ${\rm Iso}^+(\Sigma,h)$ is the Lie-group of orientation-preserving isometries of $(\Sigma,h)$.}
By setting $\alpha_\gamma(\phi(\eff)) = \phi(\eff \circ \gamma^{-1})$ for $\eff \in C_0^\infty(M,\mathbb{R})$ and
$\gamma \in G_\Sigma$, one can define the automorphisms $\alpha_\gamma : \mathcal{F}(M,g) \to \mathcal{F}(M,g)$
induced by $\gamma$ on the algebra of abstract field operators.
Then a state $\omega$ on $\mathcal{F}(M,g)$ is called {\it homogeneous} if it is invariant under the action of 
$\alpha_\gamma$, i.e.\ if
$$ \langle \alpha_\gamma(X) \rangle_\omega = \langle X \rangle_\omega \quad (\gamma \in G_\Sigma,\,X \in \mathcal{F}(M,g))\,.$$
If $\omega$ is a quasifree state, then it is homogeneous exactly if 
$$ \mathcal{W}_2^\omega(\eff_1 \circ \gamma^{-1},\eff_2 \circ \gamma^{-1}) = \mathcal{W}_2^\omega(\eff_1,\eff_2) $$
holds for all $\gamma \in G_\Sigma$ and $\eff_1,\eff_2 \in C_0^\infty(M,\mathbb{R})$.

The most important class of states for quantized linear fields on curved spacetimes, particularly in the
context of defining expectation values of the stress-energy tensor, are quasifree Hadamard states.
These are quasifree states whose 2-point functions are of so-called Hadamard form. We shall not go into 
full details of the definition of Hadamard form (see \cite{KayWald,WaldQFTCST,FewsterSmith}) for further discussion)
but mainly describe its basic features entering into the definition of the renormalized expected 
stress-energy tensor. Basically, $\mathcal{W}_2^\omega$ is of Hadamard form if
$$
  \mathcal{W}^\omega_2(\eff_1,\eff_2) = \mathcal{G}_{\rm sing}(\eff_1,\eff_2) + 
   \int \int \eff_1(x) R_\omega(x,x') \eff_2(x')\,d\mu_g(x) \, d\mu_g(x')
   $$
where $\mathcal{G}_{\rm sing}$ is a singular contribution (a distribution in 
$(C_0^\infty(M \times M))'$) which depends in a certain, local way on the spacetime metric 
$g$ and the mass parameter $m$ in the Klein-Gordon equation, but is independent of $\omega$
(i.e.\ the singular part is the same for all Hadamard states $\omega$). There remains a smooth
contribution, expressed by $R_\omega \in C^\infty(M \times M,\mathbb{C})$, which contains the dependence
on the states $\omega$. The circumstance that the singular part
$\mathcal{G}_{\rm sing}$ is the same for all Hadamard states $\omega$ is instrumental for the
definition of the expectation value $\langle : {\bf T}_{\mu\nu} : \rangle_\omega$ of the renormalized stress-energy tensor in the state $\omega$.

We shall very briefly elaborate on the ``symmetric Hadamard parametrix'' (SHP) renormalization of
the expectation value of the stress-energy tensor which was employed in \cite{SchlemmerVerch},
see also \cite{FewsterSmith}. Define the {\it symmetric Hadamard parametrix}
$$ \tilde{\mathcal{G}}(\eff_1,\eff_2) = \frac{1}{2}(\mathcal{G}_{\rm sing}(\eff_1,\eff_2) + \mathcal{G}_{\rm sing}(\eff_2,\eff_2))
+ \frac{i}{2}\mathcal{E}(\eff_1,\eff_2) $$
and, correspondingly, set 
$$ \mathcal{R}^{\rm SHP}_\omega(\eff_1,\eff_2) = \mathcal{W}^\omega_2(\eff_1,\eff_2) - \tilde{\mathcal{G}}(\eff_1,\eff_2)$$
for $\eff_1,\eff_2 \in C_0^\infty(M,\mathbb{R})$. Then the distribution $\mathcal{R}_\omega^{\rm SHP}$ is
given by a smooth integral kernel $R_\omega^{\rm SHP}$,
$$ \mathcal{R}^{\rm SHP}_\omega(\eff_1,\eff_2) = \int \int \eff_1(y) R_\omega^{\rm SHP}(x,x') \eff_2(x')\,d\mu_g(x) \, d\mu_g(x')\,.$$
Now let $x \in M$ be given and suppose that $x' \in M$ lies in a convex normal neighbourhood of $x$,
and denote by $Y^{\nu'}{}_{\nu}(x';x)$ the operation of parallelly transporting a covector
$\xi'_{\nu'}$ at $x'$ to a covector $\xi_\nu$ at $x$. Note that $Y^{\nu'}{}_{\nu}(x';x)$ depends
smoothly on $x$ and $x'$. With this convention, one can define the SHP-renormalized expectation value
of the stress-energy tensor in the state $\omega$ as 
\begin{eqnarray*}
 \langle {\bf T}^{\rm SHP}_{\mu\nu}(x)\rangle_\omega
 & = & \lim_{x' \to x}\ (\nabla_\mu \nabla_{\nu'} R^{\rm SHP}_\omega(x,x'))Y^{\nu'}{}_{\nu}(x';x) \\
 & & +\ \frac{1}{2}g^{\mu\nu}(x)(\nabla_\mu \nabla_{\nu'} R^{\rm SHP}_\omega(x,x'))Y^{\nu'}{}_{\nu}(x';x) \\
 & & +\ \frac{1}{2}R^{\rm SHP}_\omega(x,x')\,.
\end{eqnarray*}
Note that on the right hand side, $\nabla_\mu$ operates with respect to $x$ and $\nabla_{\nu'}$
operates with respect to $x'$. The resulting $\langle {\bf T}^{\rm SHP}_{\mu\nu}(x)\rangle_\omega$
is a smooth, symmetric co-tensor field of $x \in M$. This renormalization procedure has the advantage of being completely independent of any ``reference'' state; in fact, it renders a local, generally 
covariant quantity in the sense of \cite{WaldQFTCST,BFV,HollandsWaldAx}. However, in general it will
have the defect of not being divergence-free. As pointed out in \cite{Wald78,WaldQFTCST}, this defect
can be repaired as follows: One can show that there is a smooth function $Q$ on $M$, determined
entirely by the local geometry of $(M,g)$, such that 
$$ \nabla^{\mu} \langle {\bf T}^{\rm SHP}_{\mu \nu}(x) \rangle_\omega = \nabla_\nu Q(x) \quad (x \in M) $$
holds for all quasifree Hadamard states $\omega$ -- while $Q$ is state-independent. Hence, if
$Q(x)g_{\mu\nu}(x)$ is subtracted from $\langle {\bf T}^{\rm SHP}_{\mu\nu}(x)\rangle_\omega$,
the resulting quantity has vanishing divergence. In fact, this resulting quantity is a local, generally
covariant and divergence-free definition of the expectation value of the stress-energy tensor of
a quantized field which thus complies with the requirements delineated by Wald for stress-energy observables
of quantum fields
\cite{Wald78,WaldQFTCST}. Then there remains a renormalization ambiguity for this quantity to the effect
that one may add other symmetric co-tensor fields $C_{\mu \nu}$ having vanishing divergence
and determined by the local geometry of $(M,g)$. Following \cite{FewsterSmith}, we adopt the point
of view that the specification of $C_{\mu \nu}$ is a further datum (akin to the mass parameter $m$)
of the quantum field $\phi$. Using an argument of \cite{WaldQFTCST}, the freedom of choosing
$C_{\mu \nu}$ can be further narrowed down, so that 
$$ C_{\mu\nu} = C_{\mu\nu}[A,B,\Gamma,D] = A g_{\mu\nu} + B G_{\mu\nu} + \Gamma
\frac{\delta}{\delta g^{\mu \nu}}S_1(g) + D \frac{\delta}{\delta g^{\mu\nu}} S_2(g) \,,$$
where $G_{\mu\nu}$ is the Einstein tensor, $S_1(g) = \int_M R^2\,d\mu_g$, $S_2(g) = \int_M R_{\mu\nu}R^{\mu\nu}
\,d\mu_g$ with $R$ and $R_{\mu\nu}$ denoting scalar curvature and Ricci tensor, while
$\delta/\delta g^{\mu\nu}$ denotes functional differentiation; the real constants $A,B,\Gamma$ and $D$
parametrize the remaining renormalization ambiguity. In conclusion, the renormalized expected
stress-energy tensor in a quasifree Hadamard state $\omega$ takes the form
$$ \langle : {\bf T}_{\mu\nu}:(x)\rangle_\omega = \langle {\bf T}^{\rm SHP}_{\mu\nu}(x)\rangle_\omega
 - Q(x)g_{\mu\nu}(x) + C_{\mu\nu}[A,B,\Gamma,D](x) \quad (x \in M)\,.$$
Considering the case of an FRW spacetime with
cosmological time coordinate $t$, the renormalized 
expected energy density in a quasifree Hadamard state
$\omega$ with respect to cosmological time is
$$
 \langle : \boldsymbol{\varrho}:(t,{\bf x})\rangle_\omega
  = \langle :{\bf T}_{\mu\nu}:(t,{\bf x})\rangle_\omega
  \left(\frac{\partial}{\partial t}\right)^{\mu}
   \left( \frac{\partial}{\partial t} \right)^{\nu}
   \quad ((t,\bf{x}) \in \mathbb{R} \times \Sigma)\,.
   $$
It is worth making a few remarks here.

(i) The notation using double dots is to be understood
as signifying that $\langle :{\bf T}_{\mu\nu}:(x)\rangle_\omega$ is a renormalized quantity. It is not to be confused with the more common usage of
indicating ``normal ordering'', which refers to a 
reference state. The same applies to our notation of
$\langle : \boldsymbol{\varrho}:(x)\rangle_\omega$
for the renormalized expected energy density.

(ii) The above indicated procedure of defining the renormalized stress-energy expectation value of a linear
quantized field in Hadamard states $\omega$ applies not
only to the case of FRW spacetimes but to general, globally hyperbolic spacetimes.

(iii) It follows from the appearance of renormalization ambiguities in the definition of $\langle :{\bf T}_{\mu\nu}:(x)\rangle_\omega$ (parametrized by renormalization constants $A,B,\Gamma,D$) that there is
no intrinsic (without further considerations) prediction
of the total absolute value of the (expected) local energy density in quantum field theory on generic spacetimes
prior to fixing the renormalization ambiguities. The
way in which the fixing is done can have significant
implications in cosmological scenarios (see, e.g., \cite{DapFrePin}). In the
context of discussions of the role played by the
``vacuum energy'' as contribution to the cosmological constant it is
occasionally claimed that quantum field theory were
predicting a value of the vacuum energy which misses the observed value by 120 orders of magnitude. It is worth
pointing out that quantum field theory doesn't make
any such prediction.

\section{States of low energy (SLEs)}
Since our investigation later in this article refers to the case of spatially
compact FRW spacetime, we will from now on restrict our discussion to that case,
mostly to simplify notation. However, states of low energy have been introduced in
\cite{Olbermann} for all types of FRW spacetimes.

The spatially compact FRW spacetime has $\Sigma = S^3$ with the Riemannian metric
$h$ on $\Sigma$ derived from the embedding
$$\Sigma = \{(y^1,y^2,y^3,y^4) \in \mathbb{R}^4: \sum_{j = 1}^4 (y^j)^2 = 1\}$$ into
$\mathbb{R}^4$ equipped with the Euclidean metric. This Riemannian manifold carries the
metric-induced measure $d\mu_h$ which in turn gives rise to the Hilbert space
$L^2(\Sigma,d\mu_h)$. The Laplacian $\Delta_h$ is essentially selfadjoint on
$C^\infty(\Sigma,\mathbb{C})$ and there is an orthonormal basis 
$\{Y_{\bf k}\}$, ${\bf k} = (k,l,m)$, $k \in \mathbb{N}_0$, $l = 0,1,\ldots,k$,
$m = -l, -l +1,\ldots,l$, of $C^\infty$ functions on $\Sigma$ which are eigenvectors of 
$-\Delta_h$,
$$ -\Delta_h Y_{\bf k} = \kappa({\bf k})^2 Y_{\bf k}\,, \quad \kappa({\bf k}) = \sqrt{k(k +2)} \,.$$
On spatially compact FRW spacetimes, all homogeneous pure quasifree states can be represented 
in a particular form, and this will be the starting point for the definition of SLEs. To this end,
we quote the following result.
\begin{Thm} {\rm \bf (L\"uders and Roberts \cite{LudersRoberts}, Olbermann \cite{Olbermann})}${}$\\
\begin{itemize}
\item[(A)] Suppose that $\omega$ is a homogeneous pure quasifree state on
$\mathcal{F}(M,g)$ for a spatially compact FRW spacetime with scale factor $a(t)$ and Hubble function
$H(t)$. 

Then there exists a sequence $\{T_k\}$, $k \in \mathbb{N}_0$, of $C^\infty$ functions $T_k:
\mathbb{R} \to \mathbb{C}$ which are solutions of the differential equations
(where a dot indicates differentiation with respect to $t$ and overlining means complex conjugation) 
\begin{equation} \label{A1}
\overset{..}{T}_k(t) + 3 H(t)\dot{T}_k(t) + \omega_k^2(t) T_k(t) = 0 \quad (t \in \mathbb{R})
\end{equation}
with the time-dependent frequencies
\begin{equation}\label{A2}
\omega_k^2(t) = \kappa(k)/a^2(t) + m^2
\end{equation}
and which fulfill the condition
\begin{equation}\label{A3}
 \overline{T}_k(t)\dot{T}_k(t) - \dot{\overline{T}}_k(t)T_k(t) = i a^3(t) \quad (t \in \mathbb{R})\,,
\end{equation}
such that, in the sense of distributions, one has
\begin{equation}\label{A4}
\mathcal{W}_2^\omega((t,{\bf x}),(t',{\bf x}')) = \sum_{{\bf k} = (k,l,m)} \overline{T}_k(t)Y_{\bf k}({\bf x})
 T_k(t')\overline{Y}_{\bf k}({\bf x}')\,.
\end{equation}
Relation \eqref{A4} can equivalently be expressed by stating that the one-particle real-linear
map $K_\omega : C_0^\infty(M,\mathbb{R}) \to \mathcal{H}^{(1)}_\omega$ is given as follows:
\\[4pt]
$\mathcal{H}_\omega^{(1)} = \ell^2 =$ space of square-summable sequences $\{s_{\bf k}\}$ indexed by the
${\bf k} = (k,l,m)$,
\\[4pt]
$K_\omega({\eff}) = \{K_\omega({\eff})_{\bf k}\}$ with
\begin{equation}\label{A5}
K_\omega({\eff})_{\bf k} = \int_{\mathbb{R}} \int_{\Sigma} T_k(t)\overline{Y}_{\bf k}({\bf x}) {\eff}(t,{\bf x})
\,dt\,d\mu_h({\bf x})\,.
\end{equation}
\item[(B)]
Conversely, let $\{T_k\}$, $k \in \mathbb{N}_0$ be a sequence of smooth functions
$T_k : \mathbb{R} \to \mathbb{C}$ fulfilling \eqref{A1} and \eqref{A3}. Then the right hand side of
\eqref{A4} defines the two-point function $\mathcal{W}^\omega_2$ of a homogeneous pure quasifree state $\omega$
on $\mathcal{F}(M,g)$ for a spatially compact FRW spacetime with scale factor $a(t)$.
\end{itemize}
\end{Thm}
Thus, there is a one-to-one correspondence between the set of homogeneous pure quasifree states $\omega$ and
the set of sequences $\{T_k\}$ of smooth functions fulfilling \eqref{A1} and \eqref{A3}.\footnote{States which here are called ``homogeneous'' are called ``isotropic'' in \cite{Olbermann}.}

Let us, from now on, denote this set of sequences by $\mathscr{T}$. We shall next quote a result by 
Olbermann which ultimately introduces states of low energy. These make reference to weighting functions.
We shall say that that a function $f: \mathbb{R} \to \mathbb{R}$ is a {\it weighting function}
if there is a function $q \in C_0^\infty(\mathbb{R},\mathbb{R})$ such that $f(t) = q(t)^2$ $(t \in \mathbb{R})$.   
\begin{Thm} {\rm \bf (Olbermann \cite{Olbermann})} \\
For any weighting function $f$ and any 
$\{T_k\} \in \mathscr{T}$, define the sequence of numbers
\begin{equation}\label{tauf}
 \tau_j(f,\{T_k\}) = \int_{\mathbb{R}} f(t) ( |\dot{T}_j(t)|^2 + \omega_j^2(t) |T_j(t)|^2)\, dt
\quad (j \in \mathbb{N}_0)\,.
\end{equation}
\begin{itemize}
\item[$(i)$] Suppose that, for fixed weighting function $f$, the sequence $\{\overset{\circ}{T}_k\} \in 
\mathscr{T}$ has the minimizing property
\begin{equation} \label{i1}
\tau_j(f,\{\overset{\circ}{T}_k\}) = \min_{\{T_k\}\in \mathscr{T}}\, \tau_j(f,\{T_k\})
\end{equation}
for all $j \in \mathbb{N}_0$. Then the corresponding homogeneous pure quasifree state
$\overset{\circ}{\omega} = \overset{\circ}{\omega}_f$ on $\mathcal{F}(M,g)$ is a Hadamard state.
Furthermore, this homogeneous pure quasifree Hadamard state minimizes the energy density weighted
by $f$ along cosmological time, i.e.
\begin{equation}\label{i2}
\varrho_{\overset{\circ}{\omega}}[f] = \min_{\omega}\,\varrho_\omega[f]
\end{equation}
where the minimum is taken over all homogeneous pure quasifree Hadamard states  $\omega$, with
\begin{equation}\label{i3}
\varrho_\omega[f] = \int_{\mathbb{R}} f(t)\langle : \boldsymbol{\varrho}:(t,{\bf x})\rangle_\omega\,dt
\end{equation}
for any ${\bf x} \in \Sigma$. (Owing to homogenity of $\omega$, the quantity on the right hand side of
\eqref{i3} is independent of ${\bf x}$.)
\item[$(ii)$] 
For given weighting function $f$, a minimizing sequence $\{\overset{\circ}{T}_k\} \in \mathscr{T}$
exits. This sequence is unique up to choice of a phase factor for each $k$, i.e.\ 
$\{\overset{\circ}{T}_k\}$ has the minimizing property exactly if this is the case
for $\{{\rm e}^{ip_k}\overset{\circ}{T}_k\}$ for any sequence of real numbers $\{p_k\}$. All choices of
phases lead to the same minimizing state $\overset{\circ}{\omega}$.

 Given any
sequence $\{S_k\} \in \mathscr{T}$, a minimizing sequence $\{\overset{\circ}{T}_k\}$ can be
constructed from $\{S_k\}$ in the following way: One sets
\begin{equation}\label{ii1}
\overset{\circ}{T}_k(t) = \lambda(k) S_k(t) + \mu(k)\overline{S}_k(t)\,,
\end{equation}
with the definitions
\begin{align}\label{zustand}
\begin{split}
\mu(k)&=\sqrt{\frac{c_1(k)}{2\sqrt{c_1(k)^2-|c_2(k)|^2}}-\frac{1}{2}}\\
\lambda(k)&=e^{i\xi_k}\sqrt{\frac{c_1(k)}{2\sqrt{c_1(k)^2-|c_2(k)|^2}}+\frac{1}{2}}\\
c_1(k)&=\frac{1}{2}\int f(t)(|\dot{S}_k(t)|^2+\omega_{k}^2(t)|S_k(t)|^2)\,dt\\
c_2(k)&=\frac{1}{2}\int f(t)(\dot{S}_k(t)^2+\omega_k^2(t)S_k(t)^2)\,dt\\
\xi_k&=\pi-\mathrm{Arg}\,c_2(k)
\end{split}
\end{align}
\end{itemize}
\end{Thm}
{\it Remarks}\\[4pt]
($\alpha$) Note that $\{\overset{\circ}{T}_k\}$, respectively
$\overset{\circ}{\omega}$, minimize $\varrho_\omega[f]$
no matter how the renormalization constants $A,B,\Gamma,D$
have been chosen.\\[2pt]
($\beta$) For fixed ${\bf x} \in \Sigma$,
$\int_{\mathbb{R}} f(t)\,\langle : \boldsymbol{\varrho}:
(t,{\bf x})\rangle_\omega\, dt$
equals the $f$-weighted integral of the expected energy
density along the geodesic $t \mapsto (t,{\bf x})$
in FRW spacetime. It is known that this weighted integral
is bounded below as a functional on quasifree Hadamard
states $\omega$. This fact is a special case of a quantum
energy inequality, which has been established for the
minimally coupled scalar field by Fewster \cite{Fewster}.
\\[2pt]
($\gamma$) Note also that $\lambda(k)$ and
$\mu(k)$ and hence the minimizing sequence $\{\overset{\circ}{T}_k\}$ are
unchanged under a constant rescaling 
$f(t) \mapsto r \cdot f(t)$ ($r > 0$) of the amplitude of the weighting
function.
\section{Bogoliubov transformations and particle creation}

We continue to consider the case of spatially compact
FRW spacetime $(M,g)$ with scale factor $a(t)$ and
Hubble function $H(t)$.

For any pure homogeneous quasifree state $\omega$ on
$\mathcal{F}(M,g)$, the one-particle map
$K_\omega: C^\infty_0(M,\mathbb{R}) \to \mathcal{H}^{(1)}_\omega$  has $\mathcal{H}^{(1)}_\omega = \ell^2$, with $K_\omega(\eff)$ given by \eqref{A5};
$\ell^2$ is identified as the space of square summable 
sequences $\{s_{\bf k}\}$, ${\bf k} = (k,l,m)$.
The field operators $\Phi_\omega(\eff) =\pi_\omega(\phi(\eff))
$ in the GNS-representation of $\omega$ take the form
\begin{equation} \label{Phi}
\Phi_\omega(\eff) = a(K_\omega(\eff)) + a^+(K_\omega(\eff))
\end{equation}
by \eqref{fieldop1} and \eqref{fieldop2}.
Let us denote by $\{\delta_{{\bf k}',{\bf k}}\}$
the sequence in $\ell^2$ which takes the value $1$ exactly
if ${\bf k} = {\bf k}'$, and the value $0$ else.
Then write $a_{{\bf k}'} = a(\delta_{{\bf k}',{\bf k}})$
and $a_{{\bf k}'}^+ = a^+(\delta_{{\bf k}',{\bf k}})$.
With this notation, and writing more simply
${\bf k}$ instead of ${\bf k}'$, one may recast
\eqref{Phi} in the form
\begin{equation} \label{Phiinmodes}
\Phi_\omega(t,{\bf x})
 = \sum_{{\bf k} = (k,l,m)}
 (T_k(t)\overline{Y}_{\bf k}({\bf x})a^+_{\bf k}
  + \overline{T}_k(t)Y_{\bf k}({\bf x})a_{\bf k})\,,
\end{equation}
to be interpreted as an operator-valued distribution
(the right hand side becomes an operator in Fock space
$F_+(\mathcal{H}^{(1)}_\omega)$ upon integrating with
a test-function $\eff(t,{\bf x})$).  

It is known that for any pair of pure quasifree 
Hadamard states 
$\omega(1)$ and $\omega(2)$ on spatially compact FRW spacetime there
is a unitary operator $U : F_+(\mathcal{H}^{(1)}_{\omega(1)})
\to F_+(\mathcal{H}^{(1)}_{\omega(2)})$ such that
\begin{equation} \label{Uni}
\Phi_{\omega(2)}(\eff) = U \Phi_{\omega(2)}(\eff)U^{-1}
\end{equation}
for all $\eff \in C_0^\infty(M,\mathbb{R})$
\cite{WaldQFTCST,RVerch}.

Now let $\omega(1)$ and $\omega(2)$ be two pure homogeneous
quasifree Hadamard states and denote by $\{T^{(1)}_k\}$
and $\{T^{(2)}_k\}$ the corresponding sequences in
$\mathscr{T}$. Since $T_k^{(1)}$ and $T_k^{(2)}$
both fulfill the differential equation \eqref{A1} and the
solution space of \eqref{A1} is two-dimensional, there
are complex coefficients $\alpha_k$ and $\beta_k$
so that
$$
 T_k^{(2)}(t) = \alpha_k T^{(1)}_k(t) + \beta_k \overline{T}^{(1)}_k(t) \quad (t \in \mathbb{R})\,.$$
The requirement \eqref{A3} implies
$$ |\alpha_k|^2 - |\beta_k|^2 = 1\,.$$
Insertion in to \eqref{Phiinmodes} then yields the relations
\begin{eqnarray*}
\Phi_{\omega(1)}(t,{\bf x})
& = &\sum_{{\bf k} = (k,l,m)}
 (T_k^{(1)}(t)\overline{Y}_{\bf k}({\bf x})a^+_{\bf k}
  + \overline{T}_k^{(1)}(t)Y_{\bf k}({\bf x})a_{\bf k})
  \quad {\rm and} \\
\Phi_{\omega(2)}(t,{\bf x})
 &= &\sum_{{\bf k} = (k,l,m)}
 (T_k^{(2)}(t)\overline{Y}_{\bf k}({\bf x})b^+_{\bf k}
  + \overline{T}_k^{(2)}(t)Y_{\bf k}({\bf x})b_{\bf k}) 
\end{eqnarray*}
with
\begin{eqnarray}
b^+_{\bf k} &=& \alpha_k a^+_{\bf k} + \overline{\beta}_{\bf k} a_{-{\bf k}}\,, \label{Bogo1} \\
b_{\bf k} & = & \overline{\alpha}_{\bf k}a_{\bf k} + \beta_k
 a^+_{-{\bf k}}\,, \label{Bogo2} 
\end{eqnarray}
where the notation $-{\bf k} = (k,l,-m)$ for
${\bf k} = (k,l,m)$ and the property
$\overline{Y}_{\bf k} = Y_{-{\bf k}}$ have been used.  
The passage from annihilation and creation operators
$a_{\bf k}$ and $a^+_{\bf k}$ to $b_{\bf k}$ and
$b^+_{\bf k}$ by \eqref{Bogo1}, \eqref{Bogo2} is a special
case of a {\it Bogoliubov transformation}. It is related
to the unitary $U$ appearing in \eqref{Uni} as formulated
in the following, well-known theorem \cite{ShaleSt,Fulling,WaldScat,RujiAP}:
\\[4pt]
{\it
There is a unitary $U : F_+(\mathcal{H}^{(1)}_{\omega(1)})
\to F_+(\mathcal{H}^{(1)}_{\omega(2)})$ such that
$$ \Phi_{\omega(2)}(\eff) = U \Phi_{\omega(1)}(\eff)
 U^{-1} \quad (\eff \in C_0^\infty(M,\mathbb{R})) $$
exactly if 
$$ b_{\bf k} = U a_{\bf k} U^{-1} \quad \text{and}
\quad b^+_{\bf k} = U a_{\bf k}^+ U^{-1} $$
for all ${\bf k}$. This, in turn, holds if and only if
\begin{equation} \label{unimp}
 \sum_{{\bf k} = (k,l,m)} |\beta_k|^2 
= \sum_{k \in \mathbb{N}_0} (k^2 + 2k +1) |\beta_k|^2
< \infty\,. 
\end{equation} }

Now suppose that we consider the GNS-representation of
$\omega(2)$ as ``reference'' representation. With respect
to this representation, we have
$b_{\bf k} | 0\rangle_{\omega(2)} = 0$, so 
the $b_{\bf k}$ and $b_{\bf k}^+$ are annihilation and
creation operators of the spatial symmetry mode {\bf k}
with respect to this representation. One refers to this
property also by saying that $b_{\bf k}$ and $b^+_{\bf k}$
annihilate and create ``b-particles''. Correspondingly,
the number operator
$$ \boldsymbol{n}^{\omega(2)}_{\bf k} = b^+_{\bf k}b_{\bf k} $$
counts the number of of ${\bf k}$-modes with respect to
having chosen the GNS-representation of $\omega(2)$ as
reference representation. One abbreviates that by saying that $\boldsymbol{n}^{\omega(2)}_{\bf k}$ counts the
number of ${\bf k}$-modes of b-particles. Thus,
$\boldsymbol{n}^{\omega(2)}_{\bf k}$ can be regarded as a
counter for ${\bf k}$-modes which has been calibrated to
give zero response in the reference state $\langle\,.\,\rangle_{\omega(2)}$:
$$ \langle \boldsymbol{n}^{\omega(2)}_{\bf k} \rangle_{\omega(2)} = 0\,.$$
Then $\boldsymbol{n}^{\omega(2)}_{\bf k}$ gives
$|\beta_k|^2$ as response in the state $\langle\,.\,\rangle_{\omega(2)}$:
\begin{eqnarray*}
\langle \boldsymbol{n}^{\omega(2)}_{\bf k} \rangle_{\omega(1)} & = &
    {}_{\omega(1)}\langle 0| b^+_{\bf k}b_{\bf k} | 0 \rangle_{\omega(1)} \\
    & = & {}_{\omega(1)}\langle 0|
    (\alpha_k a^+_{\bf k} + \overline{\beta}_k \alpha_{-{\bf k}})
    (\overline{\alpha}_k a_{\bf k} + \beta_k a^+_{-{\bf k}})
    | 0 \rangle_{\omega(1)} \\
    & = & |\beta_k|^2
\end{eqnarray*}
observing that $a_{\bf k}|0 \rangle_{\omega(1)} = 0$
and ${}_{\omega(1)}\langle 0| a_{-{\bf k}}^+a_{-{\bf k}}
| 0 \rangle_{\omega(1)} = 1$. 

We remark that in a situation where the scale factor $a(t)$ is never constant, it is
impossible to have $|\beta_k|^2 = 0$ for all $k$ unless the weighting functions $f_1$ and
$f_2$ of the two SLEs $\omega(1)$ and $\omega(2)$ are proportional ($f_1(t) = r \cdot f_2(t)$
with a positive constant $r$). 
To see this, note that $\beta_k = 0$ implies that $b_{\bf k}|0\rangle_{\omega(1)} = 0$
which, in turn, means that $a_{\bf k} U^{-1}|0\rangle_{\omega(1)} = 0$. But the only non-zero vectors in 
$F_+(\mathcal{H}^{(1)}_{\omega(1)})$ which are annihilated by all $a_{\bf k}$ are scalar multiples of 
$|0\rangle_{\omega(1)}$. Likewise, the only non-zero vectors in $F_+(\mathcal{H}^{(1)}_{\omega(2)})$
which are annihilated by all the $b_{\bf k}$ are scalar multiples of $|0\rangle_{\omega(2)}$. Hence,
$U|0\rangle_{\omega(1)} = {\rm e}^{i r}|0\rangle_{\omega(2)}$ for some real $r$. By the properties of the
GNS-representation, this implies that $\omega(1) = \omega(2)$ as states on $\mathcal{F}(M,g)$.
However, Olbermann \cite{Olbermann} has shown that for this to hold with non-constant $a(t)$, it is
necessary that the weighting functions $f_1$ and $f_2$ are proportional. This feature for non-constant $a(t)$ 
is significantly different from the case of constant $a(t) = a_0$, where there is one unique SLE for all
weighting functions and hence always $\beta_k = 0$.

In the case of non-constant $a(t)$, the deviation of $\beta_k$ from 0 is interpreted as (cosmological) particle
creation. One may envisage the following situation portrayed in Figure 1: Some ``initial'' state $\omega(1)$ is prepared
as SLE with respect to a weighting function $f_1$ concentrated near an early cosmological time $t_1$.
Another SLE $\omega(2)$ corresponding to a weighting function $f_2$ concentrated at later cosmological
time $t_2 > t_1$ is used as reference state. This reference state can be taken as an approximate vacuum
for an observer making measurements near the time $t_2$. If the observer uses a particle (mode) counter
calibrated to give zero response in his ``approximate vacuum'' $\omega(2)$, he or she will find that
the symmetry mode ${\bf k}$ measured on the state $\omega(1)$ is excited with a distribution
$$  \langle \boldsymbol{n}^{\omega(2)}_{\bf k} \rangle_{\omega(1)} =  |\beta_k|^2 \,.$$
On the other hand, another observer making measurements around the time $t_1$ using $\omega(1)$
as reference state will find no particle excitations when making measurements on the state $\omega(1)$.
The way in which this ``particle creation'' depends on reference states that are defined with respect to
properties localized in time should be noticed here. 
\begin{center}
\epsfig{file=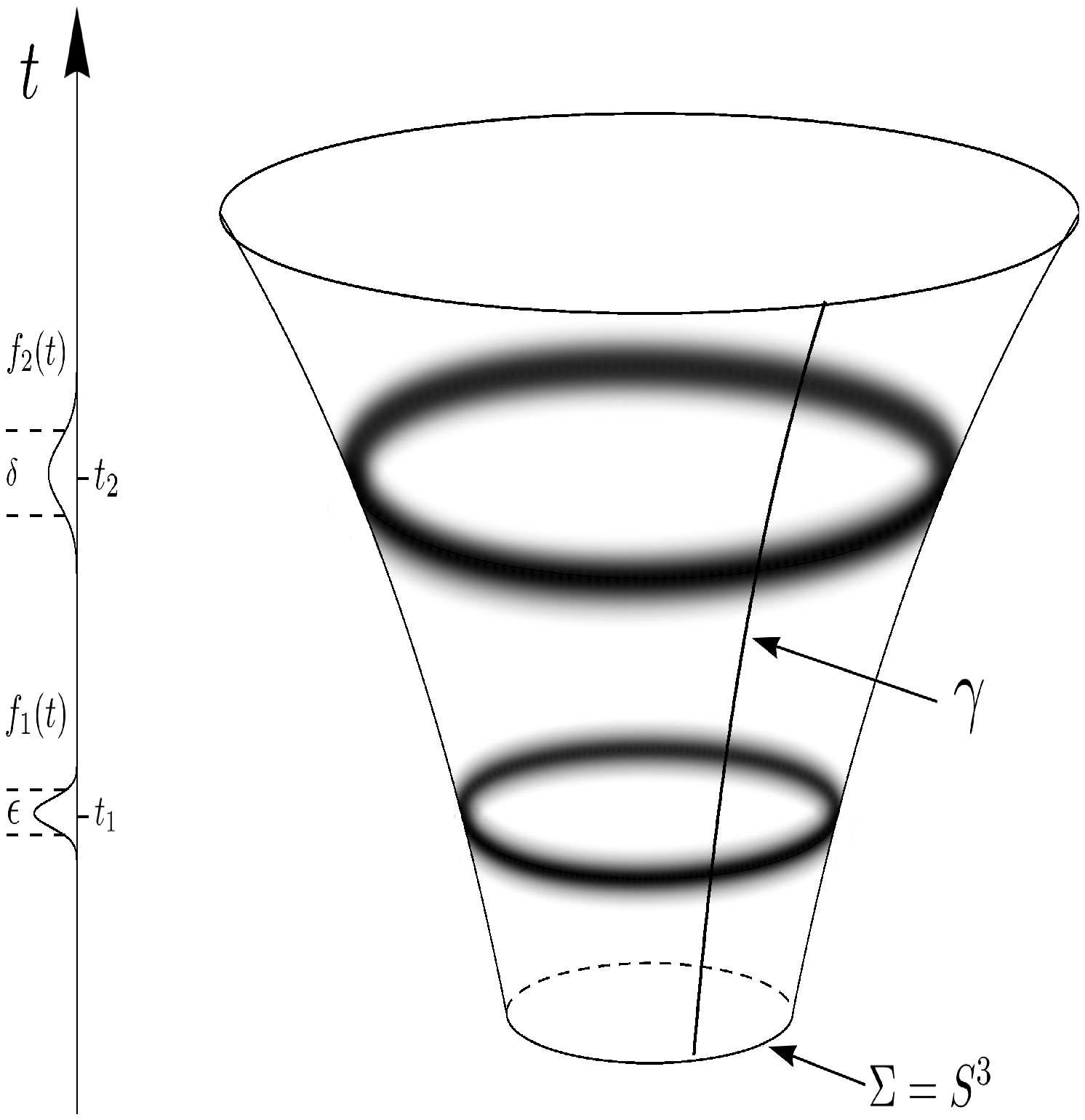,height=10cm}
\end{center}
Figure 1. {\it Illustration of the localizations of the weighting functions $f_1$ and $f_2$ in an
exponentially expanding Universe. The parameters $\delta$ and $\epsilon$
denote the characteristic widths of these functions. The curve $\gamma$ depicts the timelike geodesic
of an inertial observer.}

\section{Calculation of the number of created particles}

As described in the previous section, we want to calculate the number of
created particles per mode ${\bf k}$.
We will make some special assumptions with regard to the 
Hubble function $H(t)$ and the mass parameter $m$ and some
further approximations for the sole purpose of facilitating the numerical calculations.
Let us explain how we proceed in several steps.
\\[6pt]
{\bf (I)}\\
Supposing that a pair of weighting functions $f_1$ and $f_2$ has been chosen,
we denote by $\omega(1)$ and $\omega(2)$ the corresponding SLEs (pure, homogeneous,
quasifree) and by $\{ T_k^{(1)}\}$ and $\{T_k^{(2)}\}$ the corresponding sequences in
$\mathscr{T}$. They are related by the Bogoliubov coefficients $\alpha_k$ and
$\beta_k$ $(k \in \mathbb{N}_0)$ via
\begin{equation} \label{I1}
 T^{(2)}_k(t) = \alpha_k T_k^{(1)}(t) + \beta_k \overline{T}_k^{(1)}(t)
 \quad (t \in \mathbb{R},k \in \mathbb{N}_0)
\end{equation}
Using moreover (cf. \eqref{A3})
\begin{equation} \label{I2}
\overline{T}_k^{(j)}(t) \dot{T}_k^{(j)}(t) - \dot{\overline{T}}_k^{(j)} T_k^{(j)}(t) = i a(t)^{-3}
 \quad (t \in \mathbb{R}, k \in \mathbb{N}_0)
\end{equation}
where $a(t)$ is the scale factor of the underlying FRW spacetime, one finds that
\eqref{I1} and \eqref{I2} imply 
\begin{equation} \label{I3}
\beta_k = i a^3 \left. \left( \dot{T}_k^{(2)} T_k^{(1)} - \dot{T}_k^{(1)} T_k^{(2)} \right)\right|_{t^*}
\end{equation}
for any choice of $t^* \in \mathbb{R}$.
\\[6pt]
{\bf (II)} \\
The best way to find $\{ T_k^{(1)}\}$ and $\{T_k^{(2)}\}$ is to choose some 
$\{S_k\} \in \mathcal{T}$ and to calculate the coefficients $\lambda^{(j)}(k)$ and
$\mu^{(j)}(k)$ $(j = 1,2)$ as in (...) so that 
\begin{equation} \label{II1}
 T_k^{(j)}(t) = \lambda^{(j)}(k) S_k(t) + \mu^{(j)}(k) \overline{S}_k(t)\,.
\end{equation}
Observing that
\begin{equation} \label{II2}
 \overline{S}_k\dot{S}_k(t) - \dot{\overline{S}}_k(t) S_k(t) = i a(t)^{-3}\,,
\end{equation}
equations \eqref{I3}, \eqref{II1} and \eqref{II2} combine to yield
\begin{equation} \label{II3}
 \beta_k = \lambda^{(1)}(k) \mu^{(2)}(k) - \lambda^{(2)}(k) \mu^{(1)}(k)
\end{equation}
for all $k \in \mathbb{N}_0$.
\\[6pt]
{\bf (III)} \\
We will now specify the Hubble function as $H(t) = H$ with a constant $H > 0$, thus
\begin{equation}\label{skalen}
a(t)=e^{Ht} \,,
\end{equation}
and we will choose the unit of time such that $H =1$. In view of the fact that we have assumed 
our underlying FRW spacetime to be spatially closed, this scenario doesn't comply with a solution
of the Einstein equations with normal matter (ideal fluid) without cosmological constant; rather
it models an epoch of accelerated expansion of the Universe that may have taken place over a time-span such
that $H =1$.  For the purposes of illustration, we will nevertheless take that time-span here to
be the estimated age of the Universe, $t_H = 1.39 \cdot 10^{10} y$. (One may take any other (shorter)
time-span upon scaling the mass parameter $m$ accordingly to interpret our numerical results analogously,
see below.)
Writing as before $\kappa =\sqrt{k(k+2)}$ and inserting \eqref{skalen}, the $S_k$ must satisfy \eqref{A1},
which now takes the form
 \begin{equation}
 \ddot{S}_k(t)+3H\dot{S}_k(t)+(\kappa^2{\rm e}^{-2Ht}+m^2)S_k(t)=0.
\end{equation}
With the unit of time chosen so that $H =1$, these equations simplify further as
\begin{equation} \label{Skeqn}
 \ddot{S}_k(t) + 3 \dot{S}_k(t) + (\kappa^2 {\rm e}^{-2t} + m^2) S_k(t) = 0\,.
\end{equation}
\\[6pt]
{\bf (IV)} \\
For each $\kappa = \sqrt{k(k +2)}$, the differential equation \eqref{Skeqn} has two linearly independent
solutions, given by \begin{equation}
y_\kappa^{\pm\nu}(t)=e^{-\frac{3}{2}t}J_{\pm\nu}\Bigl(e^{-t}\kappa\Bigr)\ , 
\end{equation}
where the order of the Bessel functions is
\begin{equation}
\nu = \frac{\sqrt{9 - 4m_*^2}}{2}
\end{equation}
with $m_*$ the numerical value of the mass parameter $m$ in the units employed, i.e.\
$m_* = c^2 m /(\hbar H)$ when $m$ is given in cgs units.
The Bessel functions specialize in case that $\nu = \pm 1/2$ according to
\begin{align} \label{Bspecial}
 &J_{\frac{1}{2}}(x)=\sqrt{\frac{2}{\pi x}}\sin x &J_{-\frac{1}{2}}(x)=\sqrt{\frac{2}{\pi x}}\cos x
\end{align}
for real $x$,
 and since
this situation simplifies the numerical calculation of particle creation considerably, we
will assume that $m$ has been chosen such that $\nu = \pm 1/2$. This corresponds to $m_* = \sqrt{2}$,
or $m = \sqrt{2} \hbar H /c^2 \approx 3 \times 10^{-69}$kg in cgs units. This is surely a very small mass,
but again, this value is chosen for the purpose of illustration so as to make the
numerical calculations easier. With that choice of $m$ and correspondingly $m_* = \sqrt{2}$,
and the identities \eqref{Bspecial},
one obtains for each $k \in \mathbb{N}_0$ a solution
\begin{equation} \label{defSk}
 S_k(t)= A(\kappa) e^{-t}\cos\Bigl(e^{-t}\kappa\Bigr)+B(\kappa) e^{-t}\sin\Bigl(e^{-t}\kappa\Bigr)
\end{equation}
to \eqref{Skeqn}, 
where the coefficients
\begin{align}
A(\kappa)&=\dfrac{(2i+1)\sin\kappa}{2\kappa}+\frac{\cos\kappa}{2}\ \quad \text{and}\\
B(\kappa)&=\dfrac{-(2i+1)\cos\kappa}{2\kappa}+\frac{\sin\kappa}{2} \ .
\end{align}
have been chosen such that the initial conditions
\begin{align}
  &\dot{S}_k(0)=i\ , &S_k(0)=\frac{1}{2}\ ,
 \end{align}
are fulfilled. This ensures that the $S_k(t)$ defined in \eqref{defSk} fulfill the normalization
condition \eqref{II1}, and therefore give rise to a sequence $\{S_k\} \in \mathscr{T}$. 
\\[6pt]
{\bf (V)} \\
Now that we have chosen some $\{S_k\} \in \mathscr{T}$, we will calculate the Bogoliubov coefficients
as described in {\bf (II)} above. To this end, we must evaluate the integrals
$$\hspace{0.4cm}c_{1}^{(i)}(\kappa)=\frac{1}{2}\int\mathrm{d}tf_i(t)\left(|\dot{S}_k(t)|^2+\omega_k^2|S_k(t)|^2\right)$$ 
and
$$c_{2}^{(i)}(\kappa)=\frac{1}{2}\int\mathrm{d}tf_i(t)\left(\dot{S}_k(t)^2+\omega_k^2S_k(t)^2\right)\ ,$$
which depend on the smearing functions $f_i(t)$ ($i =1,2$) and $S_k(t)$. After some algebra one can write the integrands as 
\begin{align}
\begin{split}
  \hspace{-2cm}|\dot{S}_k(t)|^2+&\omega_k^2|S_k(t)|^2=\\
&\left(\frac{5}{4}+\frac{\kappa^2}{4}\right)I_1+\left(\frac{3}{8}+\frac{15}{8\kappa^2}\right)I_2-\frac{1}{2}I_3+\left(\frac{3}{8}-\frac{15}{8\kappa^2}\right)I_4\\
&+\left(\frac{\kappa}{4}-\frac{5}{4\kappa}\right)I_5+\frac{3}{4\kappa}I_6
\end{split}
\end{align}
and
 \begin{align}
  \begin{split}
   \dot{S}_k(t)^2+&\omega_k^2S_k(t)^2=\\
&\left(i-\frac{3}{4}+\frac{\kappa^2}{4}\right)I_1+\left(\frac{3}{8}-\frac{9}{8\kappa^2}+\frac{3i}{2\kappa^2}\right)I_2-\left(i+\frac{1}{2}\right)I_3\\
&+\left(\frac{3}{8}+\frac{9}{8\kappa^2}-\frac{3i}{2\kappa^2}\right)I_4+\left(\frac{3}{4\kappa}-\frac{i}{\kappa}+\frac{\kappa}{4}\right)I_5+\left(\frac{3}{4\kappa}+\frac{3i}{2\kappa}\right)I_6\ ,
  \end{split}
 \end{align} 
where the notation
\begin{align*}
I_1&=e^{-4t}&I_4&=e^{-2 t} \cos \left[\left(2-2 e^{-t}\right) \kappa \right]\\
I_2&=e^{-2t}&I_5&=e^{-3 t} \sin \left[\left(2-2 e^{-t}\right) \kappa \right]\\
I_3&=e^{-3 t} \cos \left[\left(2-2 e^{-t}\right) \kappa \right]&I_6&=e^{-2 t} \sin \left[\left(2-2 e^{-t}\right) \kappa \right]
\end{align*}
has been used.
\\[6pt]
{\bf (VI)} \\
As weighting functions $f_i(t)$ we would have to use squares of $C_0^\infty$-functions. However, since we
are eventually evaluating the integrals $c^{(i)}_1$ and $c^{(i)}_2$ numerically on a computer, it
is justified to approximate (up to machine precision) weighting functions by Gaussians. Therefore, for our
numerical calculations we choose as weighting
functions $f_i(t)$ Gaussians localised at the times $t_i$ and characteristic smearing widths $\epsilon$ and $\delta$
as illustrated in Figure 2 (cf.\ also Figure 1):
\begin{align}
 f_1(t)&=\dfrac{1}{\epsilon}e^{-(\frac{t-t_1}{\epsilon})^2}\ , &f_2(t)&= \dfrac{1}{\delta}e^{-(\frac{t-t_2}{\delta})^2}\,.
\end{align}
\begin{center}
\epsfig{file=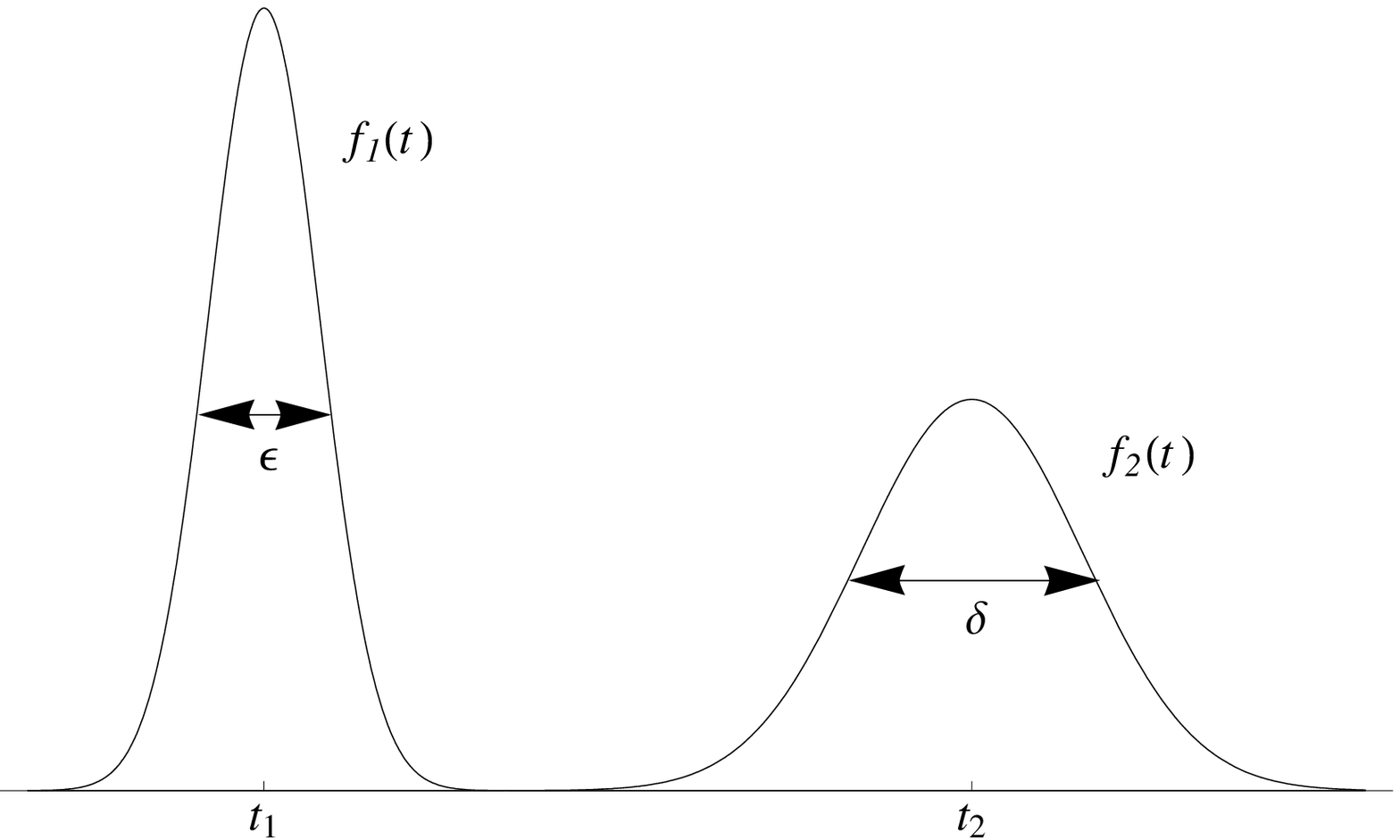,height=6cm}
\end{center}
Figure 2. {\it Illustration of the parameters $t_1,t_2,\epsilon$ and $\delta$ which characterize the weighting functions $f_1$ and $f_2$.}
\\[6pt]
As said, this is a sufficiently good approximation for a weighting function as
long as the smearing widths $\epsilon$ and $\delta$ are much smaller than the unit of time
(i.e.\ much smaller than 1 with respect to our choice of units). The normalisation factors $1/\epsilon$ and $1/\delta$ have no impact on the $\mu^{(i)}(\kappa)$ and $\lambda^{(i)}(\kappa)$ since they will cancel. Nevertheless we wrote them down for one can better imagine the limit to the Dirac distribution when $\epsilon$ and $\delta$ tend towards zero. The time coordinate is not subject to any restrictions a priori. Therefore the integrals
 $$\mathcal{I}_j(\epsilon,t_1)=\int_{-\infty}^{\infty}\mathrm{d}tf_1(t)I_j \quad (j = 1,\ldots, 6)$$ 
have to be evaluated. In order to apply the identities
\begin{align}
\int_{-\infty}^{\infty}\mathrm{d}xe^{-q^2x^2}\sin[p(x+y)]&=\dfrac{\sqrt{\pi}}{q} e^{-\frac{p^2}{4q^2}}\sin[py]\\
\int_{-\infty}^{\infty}\mathrm{d}xe^{-q^2x^2}\cos[p(x+y)]&=\dfrac{\sqrt{\pi}}{q} e^{-\frac{p^2}{4q^2}}\cos[py] 
\end{align}
for the integrations, we carry out the Taylor expansion of the argument of the trigonometric functions to first order in $t-t_1$. This approximation is justified if $\epsilon\ll 1$, since then the integrand will only be important for  $|t-t_1|\ll1$. We obtain
\begin{equation}
 2\kappa\left(1-e^{-t}\right)\approx2\kappa\left(1-e^{-t_1}\right)+\dfrac{2\kappa}{e^{t_1}}(t-t_1)\ .
\end{equation}
Multiplication of the two exponential functions yields
\begin{equation}
 e^{-(\frac{t-t_1}{\epsilon})^2}e^{-\alpha t}=e^{-\frac{1}{\epsilon^2}(t-(t_1-\frac{\alpha}{2}\epsilon^2))^2}e^{-\alpha t_1}
\end{equation}
where we  used $(\alpha \epsilon)^2\ll1$. We define a new integration variable $\tau:=t-(t_1-\frac{\alpha}{2}\epsilon^2)$ and rewrite the argument of the trigonometric function:
\begin{equation}
2\kappa(1-e^{-Ht_1})+\frac{2\kappa}{e^{t_1}}(t-t_1)=\frac{2\kappa}{e^{t_1}}\Bigl(\tau-\frac{\alpha \epsilon^2}{2}-1+e^{t_1}\Bigr) 
\end{equation}
In our case we have
\begin{align}
&q=\frac{1}{\epsilon}\ , & y &=-\frac{\alpha}{2}\epsilon^2+e^{t_1}-1\nonumber\\ 
&p=\frac{2\kappa}{e^{t_1}}\ , &py&\approx2\kappa(1-e^{-t_1})\ .
\end{align}
Consequently,
\begin{align}
 \begin{split}
  \mathcal{I}_1(\epsilon,t_1)&=\sqrt{\pi}e^{-4t_1}\\
 \mathcal{I}_2(\epsilon,t_1)&=\sqrt{\pi}e^{-2t_1}\\
\mathcal{I}_3(\epsilon,t_1)&=\sqrt{\pi}e^{-3t_1}e^{-\frac{\kappa^2\epsilon^2}{\exp(2t_1)}}\cos\left[2\kappa(1-e^{-t_1})\right]\\
\mathcal{I}_4(\epsilon,t_1)&=\sqrt{\pi}e^{-2t_1}e^{-\frac{\kappa^2\epsilon^2}{\exp(2t_1)}}\cos\left[2\kappa(1-e^{-t_1})\right]\\
\mathcal{I}_5(\epsilon,t_1)&=\sqrt{\pi}e^{-3t_1}e^{-\frac{\kappa^2\epsilon^2}{\exp(2t_1)}}\sin\left[2\kappa(1-e^{-t_1})\right]\\
\mathcal{I}_6(\epsilon,t_1)&=\sqrt{\pi}e^{-2t_1}e^{-\frac{\kappa^2\epsilon^2}{\exp(2t_1)}}\sin\left[2\kappa(1-e^{-t_1})\right]\ ,
 \end{split}
\end{align}
which completes our task of calculating the integrals $c_{1}^{(i)}$ and $c_{2}^{(i)}$
for $i = 1$. For $i=2$ the same procedure as described above applies. We just have to perform the substitutions $t_1\rightarrow t_2$ and $\epsilon\rightarrow\delta$ and to carry over the corresponding assumption $\delta\ll1$.
\\[6pt]
{\bf (VII)} \\
The explicit calculation of $|\beta_k|^2$ according to equation (\ref{II3}) ist elementary but very cumbersome. Below, we contend ourselves with calculation of  $|\beta_k|^2$ for several constellations of the parameters $\epsilon,t_1,\delta$ and $t_2$. The diagrams were generated with the program ``Mathematica''. The particle creation
coefficients $|\beta(\kappa)|^2 = |\beta_k|^2$ 
are plotted against  $\kappa=\sqrt{k(k+2)}$ continously, although actually $\kappa$ is discrete in our case where $\Sigma=S^3$. Furthermore, all times are represented in the natural unit $t_H=1,3787\times10^{10}$y.
\\[6pt]
We briefly recall our assumptions made for our numerical calculations:
\begin{itemize}
 \item The scale factor is $a(t)=e^{t}$.
\item The mass of the particle has numerical value $m_*=\sqrt{2}$ in the units used.
\item The test functions are approximated by Gaussians localised around $t_1,t_2$ and have characteristic smearing widths $\epsilon,\delta$. 
\item $\epsilon,\delta\ll1$
\end{itemize}
\subsection{The case $t_1=t_2$}
We consider the case $t_1=t_2$ and investigate the influence of $\epsilon$ and $\delta$ on the particle creation. The special case $\epsilon=\delta$ is trivial: There is no particle creation since $\omega(1)=\omega(2)$ 
\begin{center}
\epsfig{file=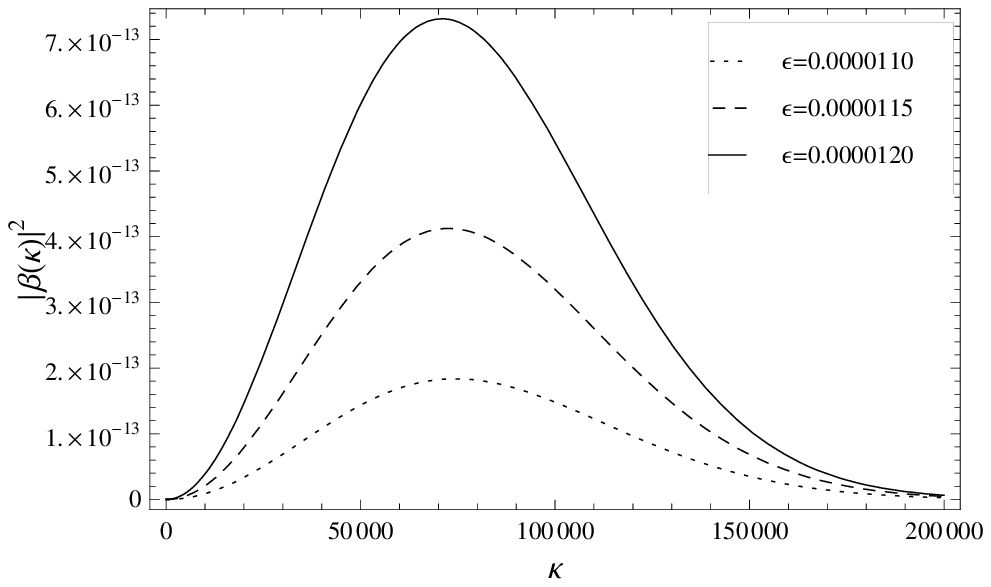,height=7cm}
\end{center}
Figure 3. {\it Particle creation rate for $t_1=t_2=0.1$, $\delta=10^{-5}$ for various values of $\epsilon$ with $\epsilon\approx\delta$}
\\[6pt]
As soon as $\epsilon$ and $\delta$ deviate minimally ($\epsilon\approx\delta$), small variations of $\epsilon$ at constant $\delta$  will affect the particle creation rate $|\beta(\kappa)|^2$ in the region $\kappa=e^{t_1}\pi/(4 \epsilon)\approx e^{t_1}\pi/(4 \delta)$ (see Figure 3). Thus the magnitude of smearing widths determines which modes will be excited.
As another case we consider a scale difference between $\epsilon$ and $\delta$ ($\epsilon\ll\delta$), cf.\ Figure 4. We remark that the maximum of the curve $|\beta(\kappa)|^2$ conforms to the larger smearing width (which is $\epsilon$ in our case) according to the formula 
\begin{equation}\label{kappamax}
\kappa_{\mathrm{max}}\approx\frac{e^{t_1}\pi}{4 \delta} 
\end{equation}
Small variations of $\epsilon$ influence $|\beta(\kappa)|^2$ only around $\kappa_{\mathrm{max}}$.
\begin{center}
\epsfig{file=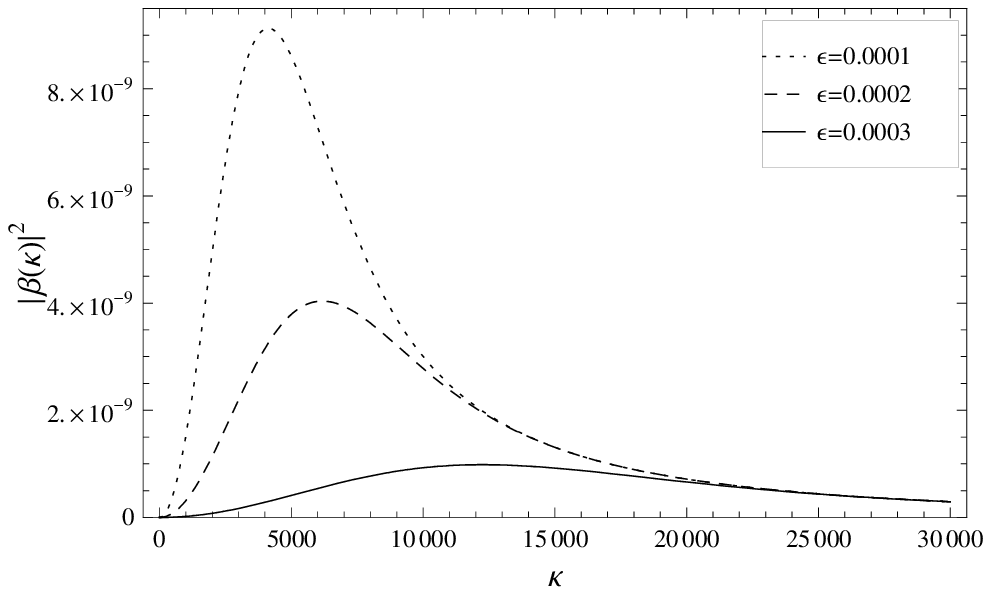,height=7cm}
\end{center}
Figure 4. {\it Particle creation rate for $t_1=t_2=0.1$, $\delta=10^{-5}$ and various values of $\epsilon$ for the case $\epsilon\gg\delta$}
${}$ \\[10pt] ${}$
\begin{center}
\epsfig{file=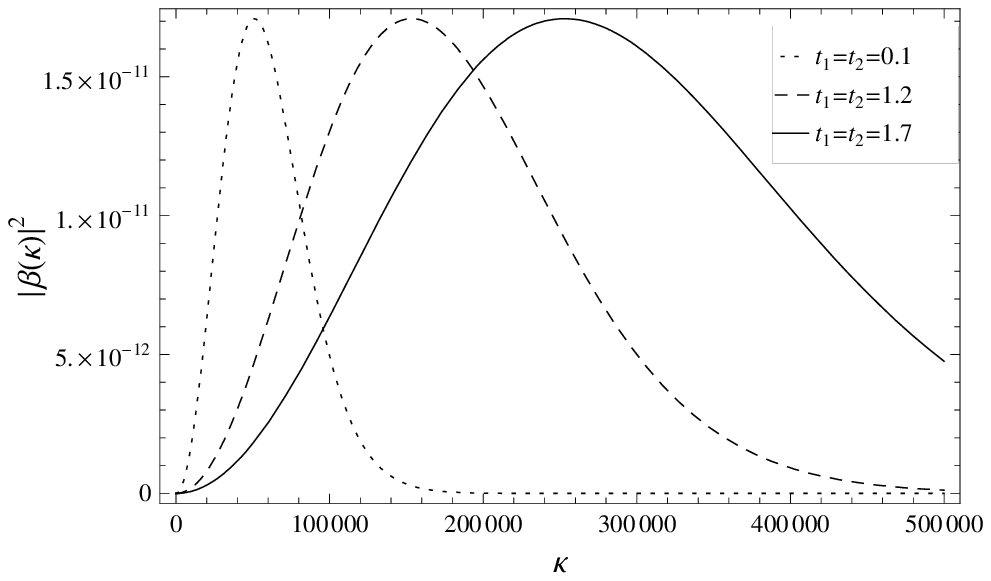,height=7cm}
\end{center}
Figure 5. {\it Particle creation rate for $\epsilon=2\delta=2\times10^{-5}$ and logarithmically increasing values of $t_1=t_2$}
\\[6pt]
Now we look at the effects caused by increasing values of $t_1=t_2$, keeping $\epsilon$ and $\delta$ fixed (Figure 5). With logarithmically growing preparation times, $\kappa_{\mathrm{max}}$ increases linearly, while $|\beta(\kappa_{\mathrm{max}})|^2$ remains constant. Simultaneously the curve broadens. This is due to the properties of the solutions $S_k(t)$ in an exponentially increasing Universe where the frequency of oscillations decays $\propto e^{-t}$. Thus the ratio of $\epsilon$ and $\delta$ (which remain constant) to the period of the oscillations decreases, leading to the same effect as shifting $\kappa_{\textrm{max}}$ to the right which would be obtained by decreasing $\epsilon$ and $\delta$ at constant $t_1,t_2$, cf.\ (\ref{kappamax}). Physically this corresponds to the redshift of a particle associated with the mode $\kappa$ which loses energy. Thus at later times the same mode will be excited more easily. In the diagram this is represented by the fact that curves corresponding to later times arise from those at earlier times by a $\kappa$-depending dilation to the right. 
%  (\ref{kappamax}) wirkt sich eine Erhöhung der Präpara\-tions\-zeiten um $\ln\Delta>0$ bei konstanten Verschmierungszeiten genauso auf die Lage von $\kappa_\textrm{max}$ aus, wie eine Verkleinerung letzterer mit dem Skalierungsfaktor $\Delta^{-1}$ bei festgehaltenen Präpara\-tions\-zeiten. Man kann nun die Verschiebung von $\kappa_\textrm{max}$ nach rechts mit wachsenden $t_1,t_2$ auch so auffassen, dass $\epsilon$ durch die zwischenzeitliche Expansion des Universums ,,geschrumpft`` ist, obwohl sich sein numerischer Wert nicht geändert hat!\footnote{Das ist die Bedeutung von ,,Schrumpfen`` im expandierenden Universum: Wenn sich alle Abstände $\propto a(t)$ vergrößern, dann scheint eine numerisch konstante Länge kleiner zu werden (und damit auch Zeiten, sofern sie durch Längen gemessen werden).} 

Finally we remark that the common feature of the case $t_1=t_2$ is $|\beta(0)|^2=0$. 

\subsection{The case $t_1\neq t_2$}
We consider now the (more general) situation when $t_1$ and $t_2$ are different. Due to the difference $t_2-t_1$ there are phase differences in the trigonometric functions entering in $|\beta(\kappa)|^2$, which manifest themselves in oscillations of the particle creation curve as seen in Figure 6. Their frequencies grow with increasing time difference $t_2-t_1$. Due to these phase differences, $|\beta(0)|^2$ is no longer equal to zero. Thus, apart from the oscillation effects there is another new remarkable feature: the particle creation effect is biggest for small modes. 
When imposing much larger time differences $t_2-t_1$, the particle creation curve grows and oscillates with higher frequency (Figure 7).
\begin{center}
 \epsfig{file=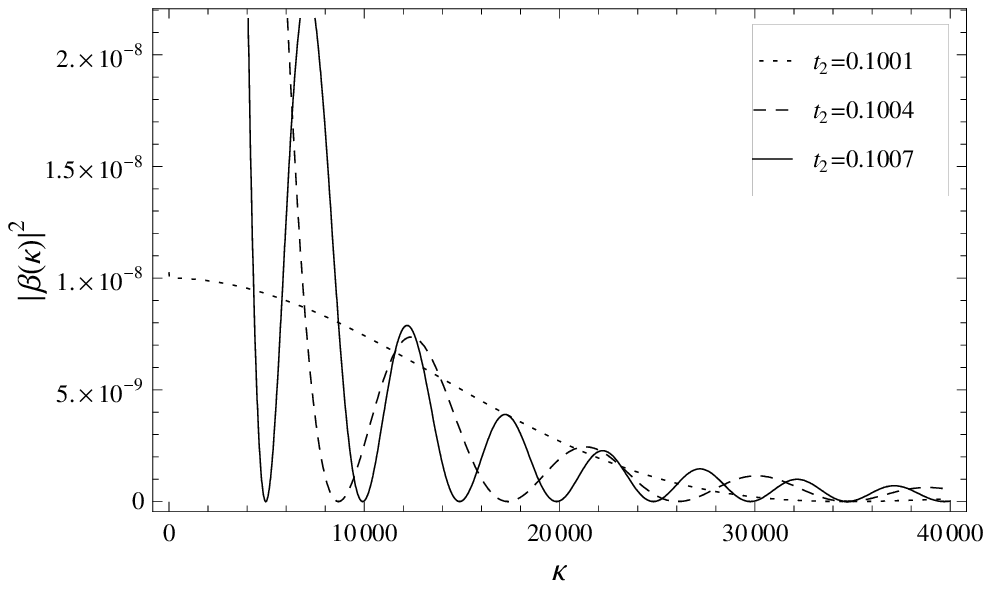,height=7cm}
\end{center}
Figure 6. {\it Particle creation rate for $\epsilon=\delta=10^{-5}$, $t_1=0.1$ and various values of $t_2$ corresponding to a small time difference}
\\[6pt]
There is some similarity with findings in \cite{HuKM}, where it is also argued that the particle number
of a state prepared at early times can diminish when compared with a ``vacuum state'' at later time. However,
the authors of \cite{HuKM} use instantaneous vacuum states as reference states, which is problematic
since they aren't (locally) unitarily equivalent to Hadamard states \cite{Fulling79,JunkerRMP}.
% Man kann sie prinzipiell messen und daraus Rückschlüsse auf $t_1$ ziehen. Allerdings wächst die Größe der Teilchenerzeugung ebenfalls mit der Differenz $t_2-t_1$ an, während die Oszillationen immer dichter werden auf der $\kappa$-Achse. Wenn man nun die charakteristischen Modulationen messen will, muss der Detektor über eine hinreichend große spektrale Auflösung verfügen. Für sehr große $t_2-t_1$ wird dann die Amplitude der Oszillationen vernachlässigbar gegenüber der Größenordnung der Teilchenerzeugung. 
\begin{center}
\epsfig{file=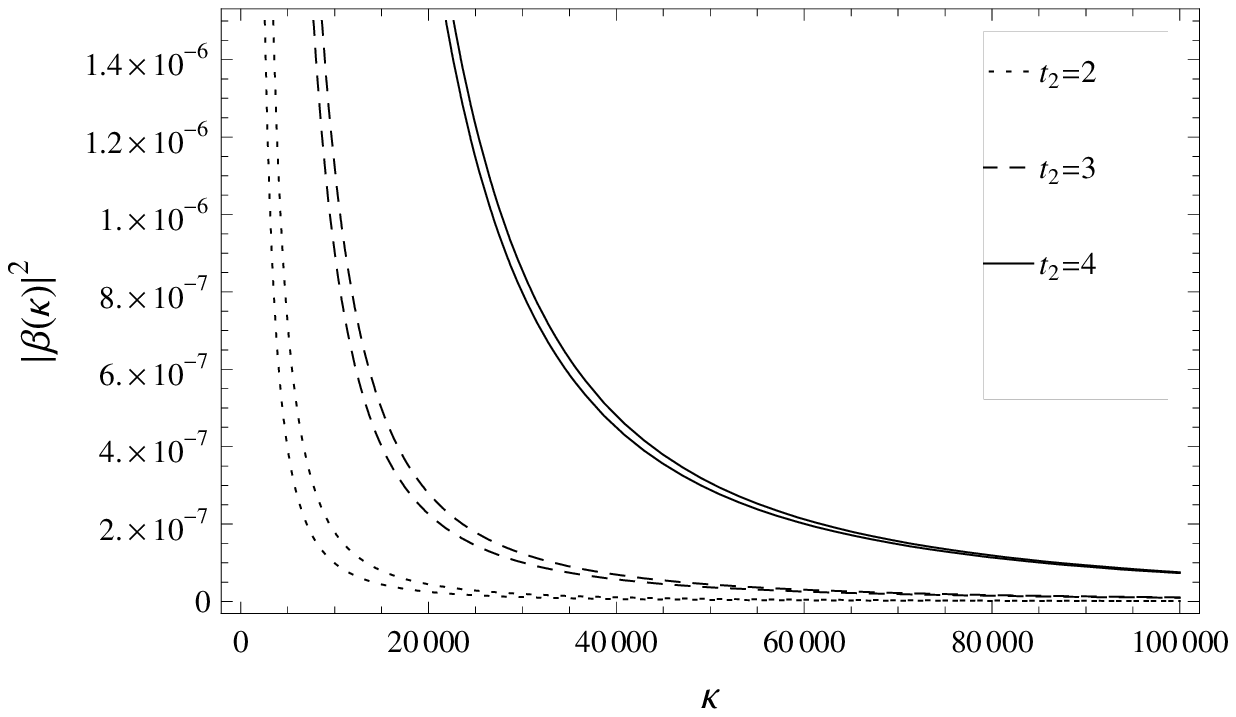,height=7cm}
\end{center}
Figure 7. {\it Envelopes of the particle creation rate for $\epsilon=\delta=10^{-5}$, $t_1=0.1$ and various values of $t_2$ corresponding to a big time difference. We have supressed the oscillations because they are of too high frequency to be properly represented. }
\\[6pt]
We have already remarked towards the end of Sec.\ 4 that in the case $t_2 > t_1$, so that $f_1$ and $f_2$ aren't proportional, it can't happen that $|\beta(\kappa)|^2 = 0$ for all values of $\kappa$. This is corroborated by the
numerical results represented in Figures 6 and 7.

Another case of interest is the limit
$\epsilon\rightarrow0$ and/or $\delta\rightarrow0$.
 In this case $f_1(t)$ and/or $f_2(t)$ would no longer be test functions, but delta distributions. 
Formally, the SLEs corresponding to the $f_i$ would, in these limits, tend to
instantaneous vacuum states, which fail to have the Hadamard property. This implies that these limiting states
are no longer unitarily equivalent to any other SLE.  This in turn should be expressed in the divergence of the total particle number in the corresponding squeezed vacuum vector, which is equivalent to $|\beta(\kappa)|^2\geq O(\kappa^{-3})$. In fact one can show (see \cite{degner}) that in our example we have for fixed $\delta>0$ in the limit $\epsilon=0$, 
$$|\beta(\kappa)|^2\geq\frac{e^{2t_1}}{4\kappa^2}\sin^2\left(2
\left(1-e^{-t_1}\right) \kappa \right)+ O(\kappa^{-3})\,.$$ 
Comparing this with the criterion \eqref{unimp}, one can see that the limiting state for $\epsilon \to 0$
isn't unitarily equivalent to the fixed SLE $\omega(2)$, and hence not a Hadamard state.

\section{Summary and outlook}

We have calculated the expected number $|\beta_k|^2$ of created particles per frequency mode
$k$ for an initial SLE at early cosmological time $t_1$ in a reference SLE at late cosmological
time $t_2$ in a closed, exponentially expanding Universe.
A characteristic feature is that $|\beta_k|^2$ shows oscillatory behaviour with respect to
variation of $k$. The envelope of $|\beta_k|^2$ decays in $k$ more strongly than $\sim k^{-3}$,
so that the oscillations are most significant for low frequency modes. The oscillatory behaviour
increases with growing time-difference $t_2 -t_1$.

A substantial drawback in discussing the possible physical significance of these findings is that they have been
obtained under considerably simplifying assumptions which cannot really be considered as physically realistic.
This refers mostly to having made the assumption of a spatially closed, exponentially expanding FRW
spacetime and having fine-tuned the mass term $m$ in the scalar field equation to an extremely small value
so as to make the numerical calculation more tractable. Above that, the quantized linear scalar field is a toy
model to which there corresponds no observed particle. It is, then, much desirable to extend the investigation
of this article to other quantum field modles such as the Dirac and electromagnetic fields, and for
physically realistic cosmological scenarios and parameters. 

However, the basic features of our methods should carry over also to these more realistic models, with
qualitatively analogous results. Therefore, the main purpose of this paper is to show that 
comological particle creation can be rigorously discussed, and detailed numerical results can be
obtained, when employing states of low energy as reference states. This, in turn, also demonstrates
the utility of the class of SLEs in the context of cosmological considerations.

\end{document}